\documentclass[10pt]{article}
\usepackage{authblk}
\usepackage{mathtools}  
\usepackage{braket}     
\usepackage{gensymb}    
\usepackage[left=1.5cm,right=1.5cm,top=1.5cm,bottom=2cm]{geometry}

\title{Polarization-driven spin precession of mesospheric sodium atoms}

\date{}

\author[1,*]{Felipe Pedreros Bustos}
\author[2]{Domenico Bonaccini Calia}
\author[1,3]{Dmitry Budker}
\author[4]{Mauro Centrone}
\author[5]{Joschua Hellemeier}
\author[5]{Paul Hickson}
\author[2]{Ronald Holzl\"ohner}
\author[6]{Simon Rochester}

\affil[1]{Johannes Gutenberg Universit\"at, Helmholtz Institute Mainz, 55128 Mainz, Germany}
\affil[2]{European Southern Observatory, D-85748 Garching b. Munich, Germany}
\affil[3]{Department of Physics, University of California, Berkeley, California 94720-7300, USA}
\affil[4]{INAF – Osservatorio Astronomico di Roma, Via Frascati, 33 00078 Monte Porzio Catone, Rome, Italy}
\affil[5]{Department of Physics and Astronomy, University of British Columbia, Vancouver, Canada}
\affil[6]{Rochester Scientific LLC, El Cerrito, California 94530, USA}
\affil[*]{Corresponding author: pedreros@uni-mainz.de}

\begin{document}

\maketitle

\begin{abstract}
We report experimental results on the first on-sky observation of atomic spin precession of mesospheric sodium driven by polarization modulation of a continuous-wave laser. The magnetic resonance was remotely detected from the ground by observing the enhancement of induced fluorescence when the driving frequency approached the precession frequency of sodium in the mesosphere, between 85~km and 100~km altitude. The experiment was performed at La Palma, and the uncertainty in the measured Larmor frequency ($\approx$260~kHz) corresponded to an error in the geomagnetic field of 0.4~mG. The results are consistent with geomagnetic field models and with the theory of light-atom interaction in the mesosphere. 
\end{abstract}

\section{Introduction}
The development of techniques for controling atomic spins have resulted in stable atomic frequency standards \cite{Vanier:2016}, high resolution spectroscopy \cite{Suefke:2017} and precise and accurate optical magnetometers \cite{Sheng:2013}. Particularly, the measurement of magnetic fields with high sensitivity using optical magnetometers has had a broad impact, from tests of fundamental physics to biomedical and geophysical applications \cite{Budker:2007}. Recently, the demonstration of optical magnetometry using naturally occurring sodium atoms in the upper mesosphere (between 85 km and 100 km altitude) opened the way for new scenarios for research in this complex and uncontrolled environment \cite{Kane:2018,PedrerosBustos:2018}. The remote detection of magnetic fields in the mesosphere brings opportunities for the study of scientific phenomena including ionic currents from oceanic tides \cite{Tyler:2003}, dynamics in the upper mantle subduction zones \cite{Blakely:2005}, and electric-current fluctuations in the ionosphere \cite{Yamazaki:2017}, among others.

The principle of an optical magnetometer is to measure the response of an atomic medium to magnetic fields using light. In particular, resonant polarized light can be employed to selectively populate energy levels in an atomic ensemble. This mechanism, called optical pumping \cite{Happer:1972}, creates polarization in both the ground and excited state of the atomic medium. The action of an external magnetic field $B$ produces precession of the atomic magnetic moments around the magnetic field at a rate given by
\begin{equation}
f_\text{Larmor} = \gamma B, \label{eq:fLarmor}
\end{equation} 
where $f_\text{Larmor}$ is the Larmor frequency, $\gamma$ is the gyromagnetic ratio of the atomic species and $B$ is the magnetic field. For sodium, the ground-state gyromagnetic ratio is $\gamma_\text{Na}=699.812$~kHz/G. When the medium is pumped at $f_\text{Larmor}$, a high degree of atomic polarization can be obtained as the atomic spins are driven synchronously with the precession during optical pumping \cite{BellBloom:1962, Alexandrov:2005}. This results in a resonance in the response of the medium whose frequency is proportional to the strength of the external magnetic field, and gives the basis for absolute and precise magnetic field measurements.  

Traditional synchronous optical pumping schemes in optical magnetometers are based on intensity- or frequency-modulated light. By contrast, we pump with polarization-modulated light. This idea was first explored for studying quantum interference in atomic systems \cite{Aleksandrov:1973} and later extended as an alternative mechanism to induce magnetic resonances in atomic magnetometers \cite{Fescenko:2013,Grujic:2013}. Recently, using a continuous-wave (CW) laser beam with polarization modulation to pump mesospheric sodium has been proposed as a mechanism for efficient optical pumping and increasing fluorescence from the sodium layer \cite{Fan:2016}. 

\begin{figure}[t]
\centering
\includegraphics[width=0.4\linewidth]{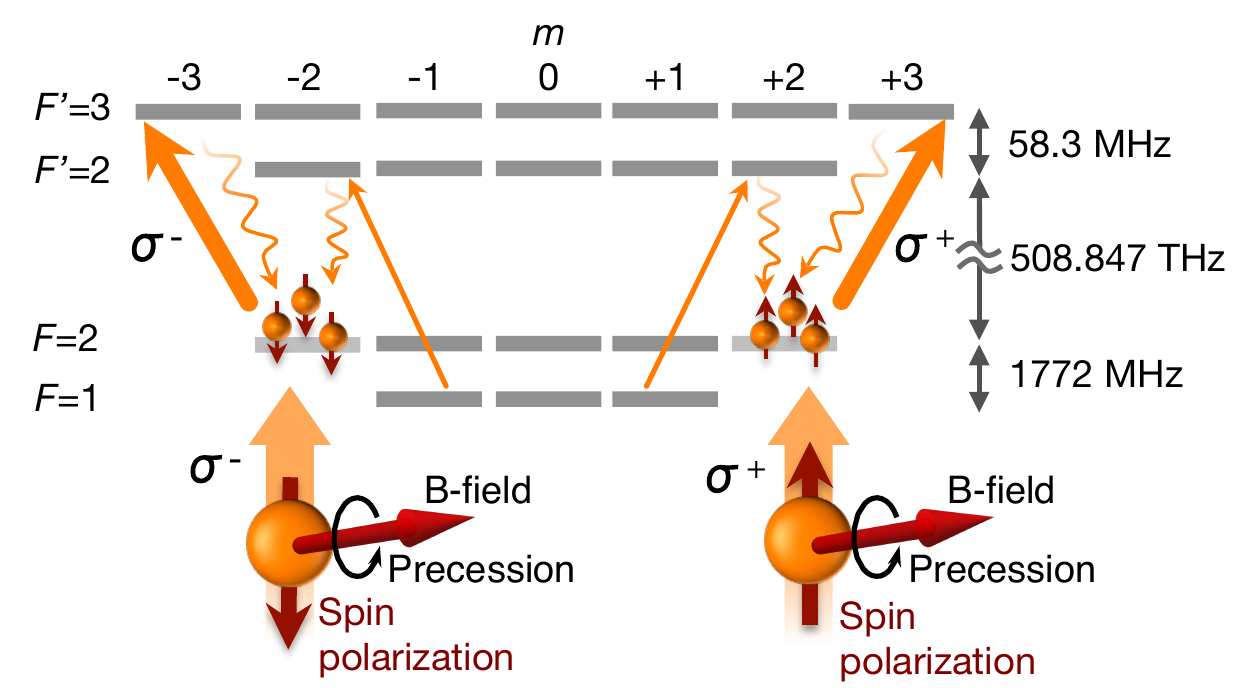}
\caption{Simplified diagram of the hyperfine structure of sodium including magnetic sublevels. The quantization axis is chosen along the direction of the light propagation. Alternating circularly polarized light (thick orange arrows) at the Larmor frequency increases spin polarization in the $\ket{F=2,m=\pm2}$ states. $\sigma^+$ denotes left-hand circular polarization and $\sigma^-$ denotes right-hand circular polarization. The thin orange arrows indicate repumping used to recover atoms lost in the $\ket{F=1}$ ground state.
}
\label{fig:levels}
\end{figure}

In this work, we demonstrate in an on-sky experiment synchronous optical pumping of mesospheric sodium using a CW polarization-modulated laser beam and report the observation of a magnetic resonance when driving at the Larmor frequency. The pumping process is schematically depicted in Fig.~\ref{fig:levels}. Photons with left-hand circular polarization ($\sigma^+$) are absorbed and drive transitions from ground to an excited state with $\Delta m=+1$, where $m$ is the magnetic quantum number. The atoms then spontaneously decay. This increases the atomic spin polarization along the light propagation direction. However, if a magnetic field is applied perpendicular to the light propagation direction, atomic spins Larmor-precess around this field. Half of the Larmor period $1/(2f_\text{Larmor})$ after pumping, the ground-state spin polarization points counter to the light propagation direction, and further pumping with left-hand circular polarization only depolarizes the medium. However, with the atomic polarization pointing in this direction, right-hand circular polarization ($\sigma^-$) can be used to increase the spin polarization; by continued polarization-switching every half-Larmor cycle, atoms can be pumped to the $|m|=F$ end-states. This increases the return flux by confining atoms to strong cycling transitions with fluorescence preferentially along the laser propagation axis. 

\section{Experiment}

\begin{figure}[t]
\centering
\includegraphics[width=0.4\linewidth]{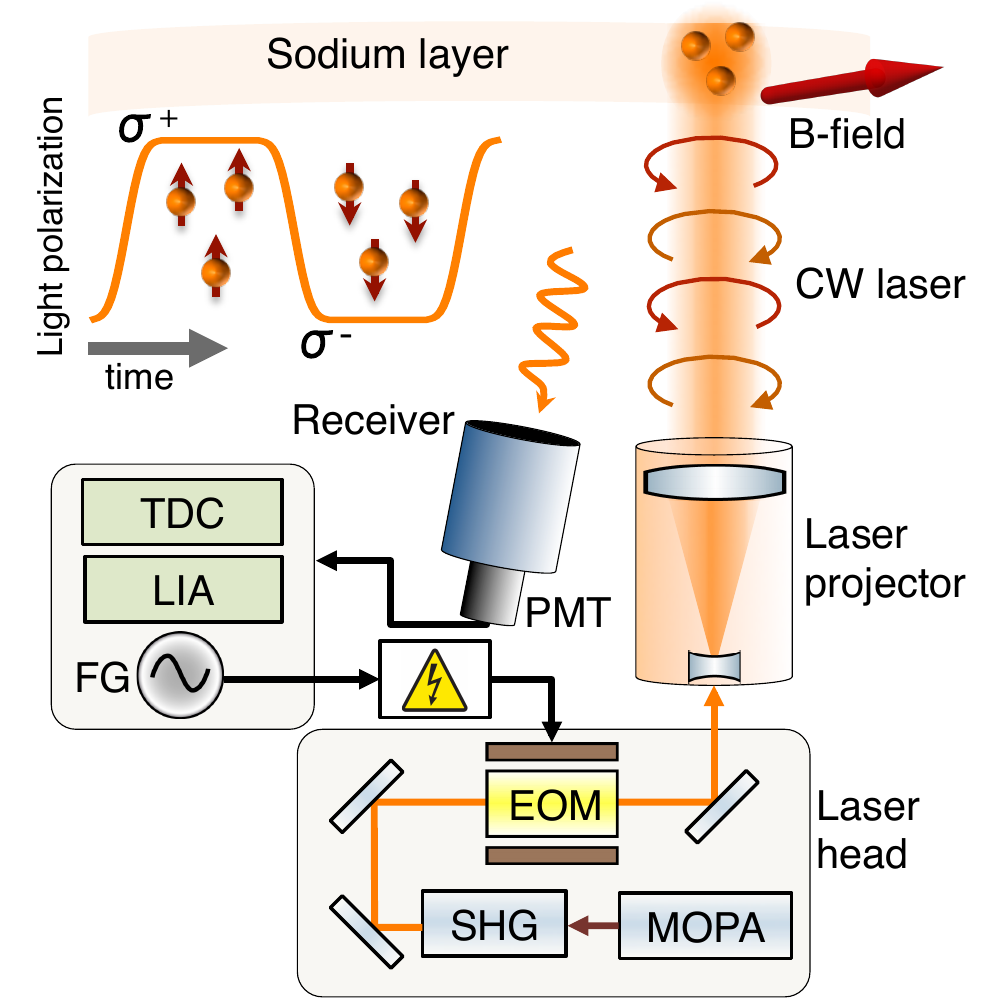}
\caption{Schematic of the experimental arrangement. MOPA: master-oscillator-power-amplifier laser, SHG: second-harmonic generation, EOM: electro-optic modulator, PMT: photomultiplier tube, TDC: time-to-digital converter, LIA: lock-in amplifier, FG: function generator. The diagram in the upper-left corner indicates the orientation of the atomic spins with the time-varying polarization of the laser beam.
}
\label{fig:setup}
\end{figure}

The experiment was carried out at the Observatorio del Roque de los Muchachos, La Palma, during April 16th--17th, 2018. The experiment is schematically presented in Fig.~\ref{fig:setup}. The laser system was built by the European Southen Observatory (ESO) as a prototype for the new generation of laser guide star devices \cite{Bonaccini:2012}. A master-oscillator-power-amplifier (MOPA) laser produces a 40 W CW beam at 1178~nm. The second-harmonic generation (SHG) crystal doubles the frequency of the infrared beam to produce a 20 W CW beam with a vacuum wavelength of 589.16~nm and a linewidth of $\approx$2~MHz. A function generator (Rigol 1062Z) followed by a power amplifier supplies a high-voltage signal to a custom-made broadband electro-optic modulator (EOM, Qubig D7v-R3-589). The electric field applied to the crystal introduces modulation of polarization of the laser beam. The laser beam is expanded through a Galilean telescope to a diameter of 30~cm. Photons emitted by the excited sodium atoms are collected on the ground with a commercial telescope of 40~cm diameter (Celestron Edge HD 1300) placed eight meters from the laser-launch telescope. A custom-made photometer consists of a pinhole placed at the focal plane of the telescope to isolate the sodium fluorescent spot, a narrowband filter of 0.30(5)~nm bandwidth centered at the resonant wavelength of sodium, and a photomultiplier tube (PMT, Hamamatsu H7422). The analog photon-detection pulses from the PMT are filtered and amplified with a discriminator (Ortec 9327, not shown) producing clean TTL (transistor-transistor logic) pulses of 100~ns width. Individual pulses are time-tagged, recorded and transferred to a computer with a fast time-to-digital converter (TDC, Roithner LaserTechnik TTM8000). A custom-made integrator amplifier (not shown) with a 1-dB bandwidth of 900~Hz converts the sequence of TTL pulses into a slowly-varying signal whose amplitude is proportional to the photon flux in the magnetic-resonance region. This signal is fed into the lock-in amplifier (LIA, Stanford Research SR865) that demodulates and extracts in real time the root-mean-square (RMS) amplitude of the AC component. With this approach, two parallel acquisition systems (TDC and LIA) provide redundancy and independent outputs for cross-checking. 

\pagebreak 
\section{Results}

\begin{figure}[t]
\centering
\includegraphics[width=0.4\linewidth]{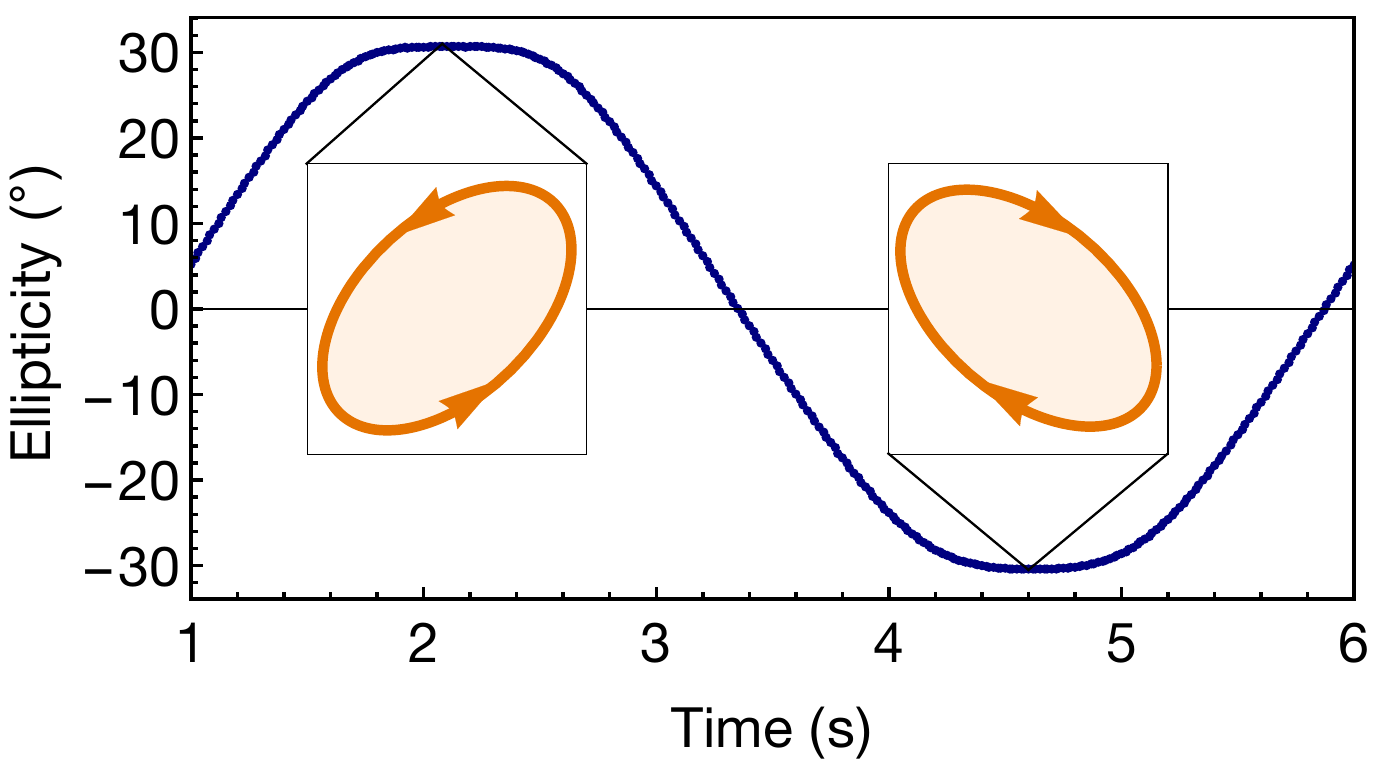}
\caption{Ellipticity of the laser beam measured at the output of the launch telescope. Here the modulation frequency is much lower than the Larmor frequency due to limitations of the polarimeter response. The polarization ellipses at the maximum and minimum ellipticity are also shown.}
\label{fig:ellipticity}
\end{figure}

The polarization of the laser beam was calibrated with a polarimeter (Thorlabs PAX5710VIS-T) placed at the output of the laser projector. A high-voltage signal of approximately $\pm$350~V was used to drive the EOM at low frequency (Fig.~\ref{fig:ellipticity}). A maximum ellipticity excursion of $\pm 30$\degree\ was achieved (the full $\sigma^+$ to $\sigma^-$ range is $\pm 45$\degree). We suspect that thermal birefringence due to heat transfer to the crystal when operating at high optical power, combined with a limited maximum voltage that can be supplied to the EOM, prevented us from reaching $\pm 45$\degree\ ellipticity in the current setup. Proper crystal thermal management as well as improvements in light polarization purity through the optical elements of the laser head will be established for future experiments. 

The use of polarization-modulated light in this approach has several advantages with respect to the intensity modulation used previously, for instance, the increase of the average photon return flux at fixed laser power, increasing the signal-to-noise ratio (SNR) for detection on the ground. The atmosphere does not impose a serious limitation regarding depolarization of the pump laser beam due to proagation through the turbulent layers, as shown by theoretical models and experiments on optical communications, where the depolarization due to the atmosphere was found to be less than 1\% \cite{Collett:1972,Toyoshima:2009}.

Scintillation due to atmospheric turbulence produces brightness fluctuations of the fluorescent sodium when seen from the surface of the Earth (a phenomenon commonly known as "star twinkling") that manifests in this experiment as detrimental intensity noise. To address the effects of scintillation, we dither the modulation frequency at a higher rate than the scintillation variations during the scan. Dithering the modulation frequency during the scan can also be beneficial in suppressing intensity fluctuations derived from thin cirrus clouds and density fluctuations in the sodium layer. The power spectrum of scintillation from stars measured at La Palma falls above 100~Hz \cite{Dravins:1998a}, thus we choose a dither frequency of $f_\text{m}=300$~Hz. The frequency of the polarization modulation is:

\begin{equation}
f_\text{pol}(t) = f_\text{center} + \delta f \cdot  \cos(2\pi f_\text{m} t), \label{eq:dither}
\end{equation}
where $f_\text{center}$ is the central frequency of the modulation and $\delta f$ is the excursion of the dither. 
For this experiment, the excursion was chosen as $\delta f=15$~kHz in order to resolve narrower magnetic resonance features predicted by numerical simulations. 

The received signal must be demodulated to extract the magnetic resonance. The demodulation was carried out with a lock-in amplifier to detect the amplitude and phase of the signal component oscillating at $f_\text{m}$ and, independently and in parallel, by calculating the ratio of photon counts within consecutive semi-cycles of the dither period. The demodulated signal shows a point symmetric dip and peak, separated by $2\delta f$, corresponding to the rising and falling slopes of the magnetic resonance probed during the frequency scan. The demodulation process produces the output  $S(f_\text{center} + \delta f) - S(f_\text{center} - \delta f)$ where $S(f_\text{center})$ is the flux at $f_\text{center}$, resulting in two opposite sign copies of the original magnetic resonance $S(f)$ separated by $2\delta f$.

\begin{figure}[t]
\centering
\includegraphics[width=0.4\linewidth]{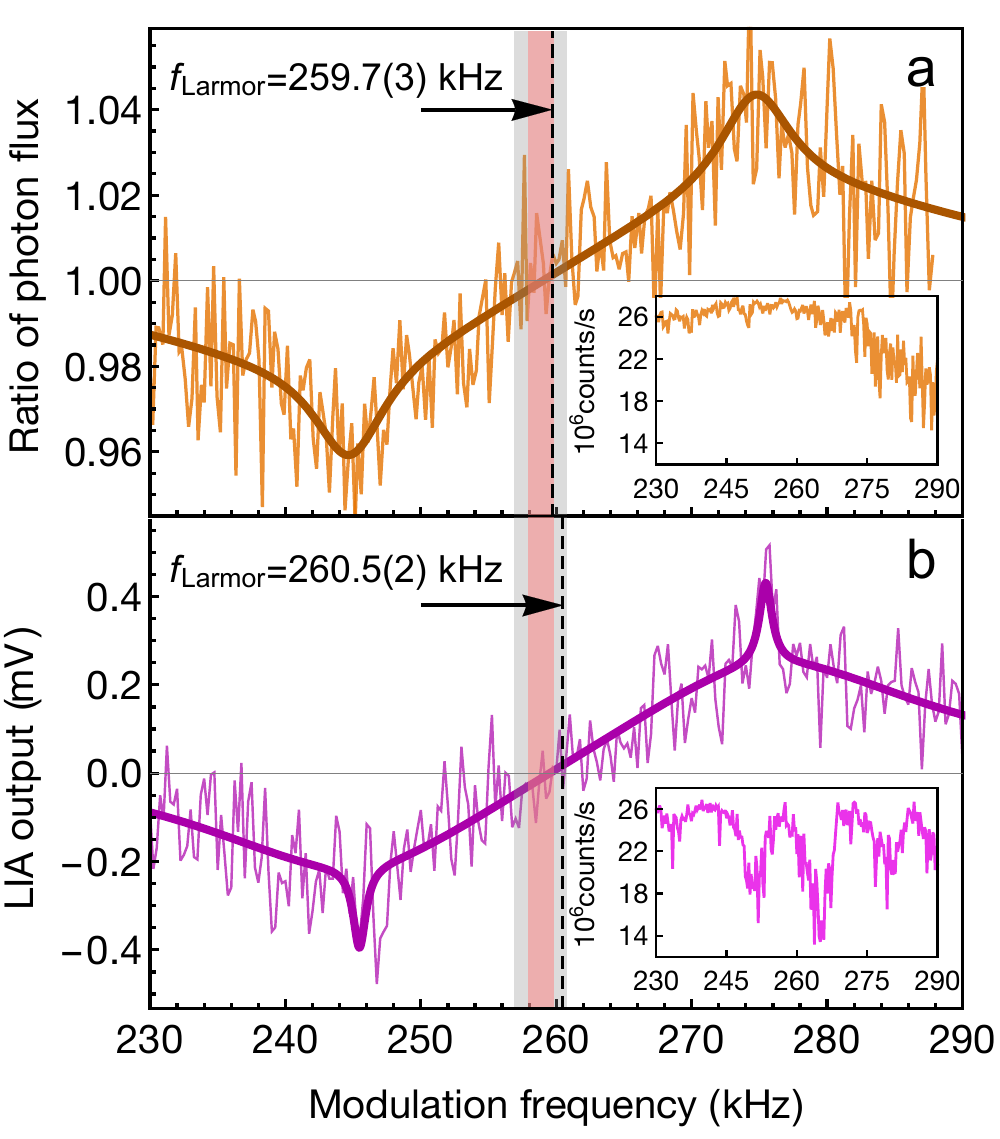}
\caption{Demodulated magnetic resonances obtained from the ratio of photon counts (a) and with lock-in amplifier (b). Each resonance consists of one peak and one dip representing two copies of the original resonance. The Larmor frequency is found in the midpoint between peaks (dashed lines). The thick line is a fit of two superimposed Lorentzians separated by $2\delta f$. The vertical red band represents the Larmor frequency corresponding to the geomagnetic field between 85~km and 100~km altitude according to the WMM2015. The gray vertical band indicates the uncertainty of the prediction from the WMM2015. The inset figures show the photon count rate during the frequency scan.}
\label{fig:results}
\end{figure}

Results of two scans performed at an elevation of 64\degree\ and azimuth of $+3$\degree\ are shown in Fig.~\ref{fig:results}. The total laser power output was 12.5 W of which 10\% was used for D$_2$b repumping (sideband at +1713~MHz from resonance). The angle between the geomagnetic field and the laser beam was $\theta_B= 80$\degree. The frequency axis shows the value of $f_\text{center}$ during the scan. Each peak is fitted with a double Lorentzian, representing broad and narrow components of the resonance. The Larmor frequency is found at the midpoint between peaks and is a free parameter of the model, in addition to the amplitude and width of the peaks. The separation of the peaks is fixed at $2\delta f$. The geomagnetic field predicted by the World Magnetic Model 2015 (WMM2015)\cite{WMM:2015} at 85~km and 100~km altitude above La Palma is 0.3713(15)~G and 0.3685(15)~G respectively, which from Eq.~\ref{eq:fLarmor} corresponds to a Larmor frequency of 259.8(1.0)~kHz and 257.9(1.0)~kHz. The estimates of the Larmor frequency obtained from fitting of the resonances shown in Fig.~\ref{fig:results} are 259.7(3)~kHz and 260.5(2)~kHz, close to the prediction of the WMM2015 at altitudes within the boundaries of the sodium layer (as also shown in Fig.~\ref{fig:results}).

The reading of our sky magnetometer can, in principle, be affected by vertical winds in the mesosphere and fast transient spikes in the sodium density within the sodium layer \cite{Pfrommer:2014} leading to atoms precessing at different Larmor frequencies due to vertical magnetic field gradients ($dB/dH=-1.83 \times 10^{-4}$~G/km at La Palma according to the WMM2015). For these reasons, parallel measurements of the vertical sodium density profile, for example, using a Light Detection and Ranging (LIDAR) system, as well as observing the fluorescence from small vertical layers of the sodium profile, could help to increase the accuracy of the Larmor frequency measurement in the vertical direction.

The resonance of Fig.~\ref{fig:results}.a shows a flux enhancement of $\sim$4\% at the Larmor frequency with respect to the flux at a modulation frequency of 230~kHz. The total photon-count rate during the scan is shown in the inset figure for each magnetic resonance in Fig.~\ref{fig:results}. The strong fluctuations in the photon flux during the scan are a result of atmospheric turbulence and the sporadic presence of variable thin layers of dust in the troposphere due to the Saharan Air Layer passing over the Canary Islands. The reduction of the photon flux towards the end of the scan (inset Fig.~\ref{fig:results}.a) and the oscillating flux during the scan (inset Fig.~\ref{fig:results}.b) are a consequence of tracking errors in the receiver telescope which resulted in the fluorescent spot not being perfectly centered in the pinhole prior to the photomultiplier. The noise seen in the magnetic resonance is primarily due to shot noise and the residual effect of high-frequency scintillation. 
 
\begin{figure}[t]
\centering
\includegraphics[width=0.4\linewidth]{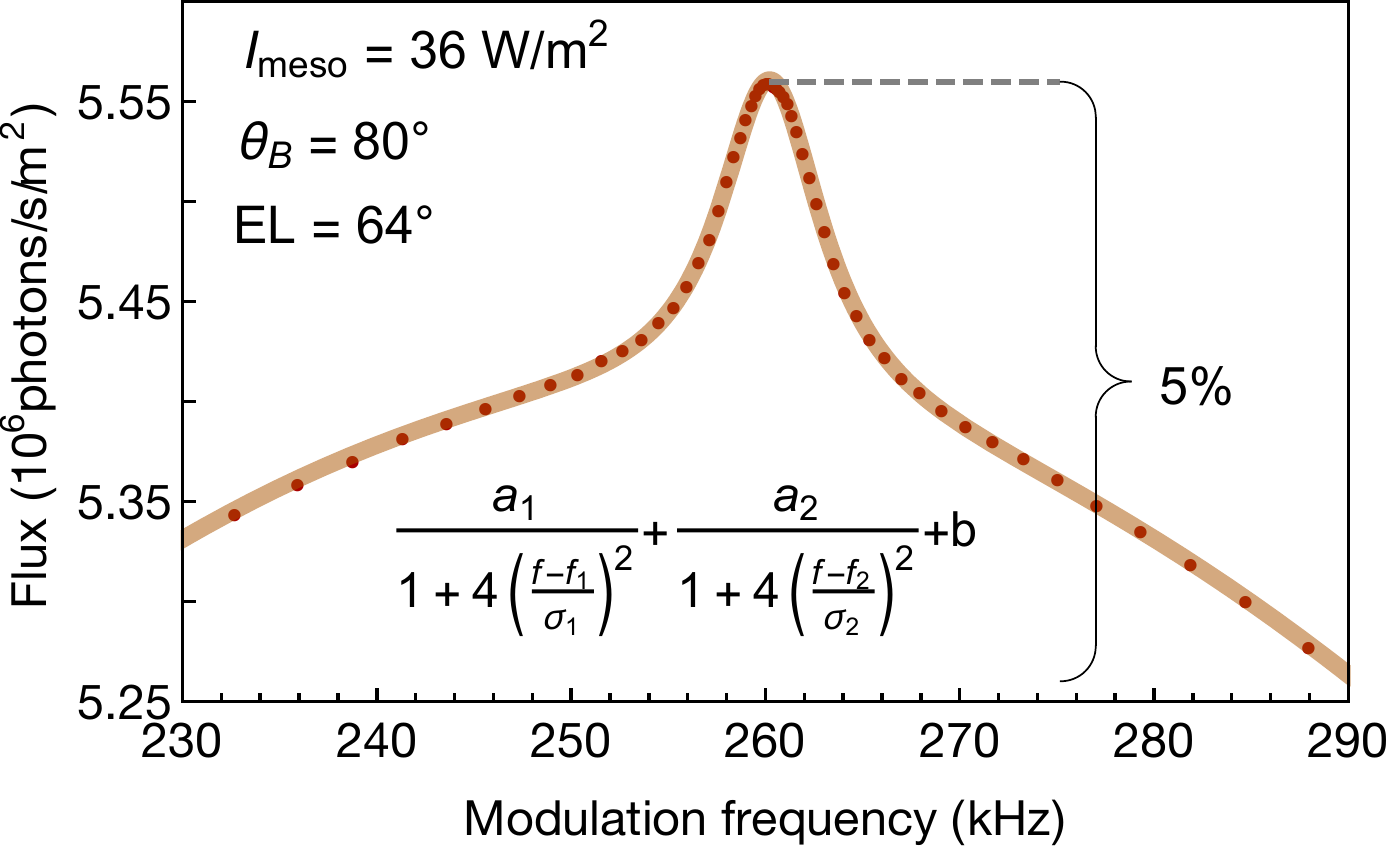}
\caption{Numerical simulations of the magnetic resonance (dots) and a double Lorentzian model (line), where $a_{1,2}$ are the amplitudes, $f_{1,2}$ are the central frequencies, $\sigma_{1,2}$ are the full-width-half-maximum of each Lorentzian and $b$ is an offset.}
\label{fig:simulation}
\end{figure}

Numerical simulations of the time evolution of the sodium atomic polarization using a density-matrix model \cite{Holz:2010a} show that the measured signal in our setup can be well described by two superposed Lorentzians (Fig.~\ref{fig:simulation}). This observation supports our choice of fit functions of the magnetic resonances shown in Fig.~\ref{fig:results}. In addition, the return-flux enhancement found from modeling shows good agreement with the data using the estimated experimental conditions for the measurements. The parameters used to simulate the magnetic resonance shown in Fig.~\ref{fig:simulation} are given in Table~\ref{table:parameters}.

Simulations also show that the broad Lorentzian resonance is centered at a slightly different frequency off the narrow resonance at high irradiance. This effect is due to light shift due to real transitions \cite{Bulos:1971} in which atoms precess in the excited state at a different frequency compared to that of ground-state; however, we expect to observe the shift in future measurements with better signal-to-noise ratio.

\renewcommand{\arraystretch}{1}

\begin{table}[htbp]
\centering
\caption{Simulation parameters with the corresponding symbols and values.}
\begin{tabular}{lc}
\hline
Beam transit rate ($\gamma_\text{beam}$) & 1/(5.0 ms) \\ 
Velocity-changing collisions rate ($\gamma_\text{vcc}$) & 1/(15 $\mu$s) \\ 
Spin-damping collisions rate ($\gamma_\text{sd}$) & 1/(150 $\mu$s) \\ 
Mesospheric temperature ($T_\text{meso}$) & 200 K \\
Laser irradiance at the mesosphere ($I_\text{meso}$)& 36 W/m$^2$\\  
Laser linewidth ($\Delta f$) & 2 MHz \\ 
D$_2$b repumping frequency offset ($f_{\text{D}_2\text{b}}$) & 1713 MHz \\
Repumping fraction ($q$) 		& 10\% \\ 
Gyromagnetic ratio of sodium ($\gamma_\text{Na}$) & 699.812 kHz/G \\ 
Magnetic field strength ($B$) & 0.371 G \\
Magnetic polar angle ($\theta_B$) & 80\degree\ \\
\hline
\end{tabular}
\label{table:parameters}
\end{table}

\pagebreak
 
\section{Conclusions}
 
In conclusion, we have observed for the first time, to the best of our knowledge, the atomic spin precession of mesospheric sodium induced by modulating the polarization of resonant continuous-wave laser beam at 589.16~nm. Severe weather conditions at La Palma allowed only two nights of observations. Therefore, additional measurement campaigns are foreseen to carry out a more complete study of the parameter space in order to understand the merits and shortcomings of this approach. Immediate possible applications include the mapping of geomagnetic fields at mesospheric scales and boosting the brightness of sodium laser guide stars when pointing in directions with large angles between the laser beam and the geomagnetic field lines.\\

\noindent {\sffamily\bfseries{Funding.}} Office of Naval Research Global (ONRG) (N62909-16-1-2113); 
Natural Sciences and Engineering Research Council of Canada and
Canada Foundation for Innovation.\\

\noindent {\sffamily\bfseries{Acknowledgements.}} F.P.B. acknowledges the support of a doctoral scholarship from the Carl-Zeiss Foundation. The authors thank the Isaac Newton Group (UK) and the Instituto Astrofisico de Canarias at the Observatorio del Roque de los Muchachos for their support during the measurements campaign. The reported observations were carried out in the frame of ESO Laser Systems R\&D activities.


\end{document}